\documentstyle[aps,eqsecnum,epsf,tighten]{revtex}
\def\ltsima{$\; \buildrel < \over \sim \;$}
\def\simlt{\lower.5ex\hbox{\ltsima}}
\def\gtsima{$\; \buildrel > \over \sim \;$}
\def\simgt{\lower.5ex\hbox{\gtsima}}
\newcommand{\be}{\begin{equation}}
\newcommand{\ee}{\end{equation}}
\newcommand{\vtheta}{\mbox{\boldmath $\theta$}}
\newcommand{\vphi}{\mbox{\boldmath $\phi$}}
\newcommand{\vdelta}{\mbox{\boldmath $\delta$}}
\newcommand{\vepsilon}{\mbox{\boldmath $\epsilon$}}
\newcommand{\trp}{{\mbox{\footnotesize{\sf T}}}}
\newcommand{\COCCIAGWE}{{\em Gravitational Wave Experiments},
edited by E. Coccia, G. Pizzella and F. Ronga 
(World Scientific, Singapore, 1995)}
\newcommand{\SCHUTZJF}{{\em Gravitational Wave Data Analysis}, edited by
B.F. Schutz (Kluwer,  Dordrecht, 1989)}
\draft
\title{Bayesian Bounds on Parameter Estimation Accuracy for 
       Compact Coalescing Binary Gravitational Wave Signals}
\author{David Nicholson and Alberto Vecchio\footnote{Present address:
Max Planck Institut f\"{u}r Gravitationsphysik, Albert Einstein Institut,
Schlaazweg 1, D-14473 Potsdam.}}
\address{Department of Physics and Astronomy,\\
University of Wales College of Cardiff, PO Box 913, Cardiff CF2 3YB, UK.}
\date{\today}
\begin{document}
\maketitle
\begin{abstract}
A global network of very sensitive large-scale laser 
interferometric gravitational wave detectors is projected 
to be in operation by around the turn of the century. 
The network is anticipated to bring a range of new 
astrophysical information --- relating to neutron stars, 
black holes, and the very early universe  --- and also
new fundamental physics information, relating to the nature 
of gravity in the strongly nonlinear regime for example. This 
information is borne by gravitational waves that will 
typically be very much weaker than the level of intrinsic
strain noise in the detectors. Sophisticated signal 
extraction methods will therefore be required to analyze
the network's data. Here, the noisy output of a 
{\em single} laser interferometric detector is examined.
A gravitational wave is assumed to have been detected in
the data. This paper is concerned only with 
the subsequent problem of parameter estimation.  
Specifically, we investigate theoretical lower 
bounds on the minimum mean-square errors (MSE) associated 
with measuring the parameters that characterize the waveform.
The pre-merger inspiral waveform generated by an orbiting
system of neutron stars or black holes is ideal for this 
study. Monte Carlo measurements of the parameters of noisy
inspiral waveforms have been performed
elsewhere, and the results must now confront statistical signal 
processing theory.
Three theoretical lower bounds on parameter estimation 
accuracy are considered here: the Cramer-Rao bound (CRB); 
the Weiss-Weinstein bound (WWB); and the Ziv-Zakai bound (ZZB). 
The CRB is the simplest and most well-known of these bounds, 
but suffers from a number 
of limitations. It has been applied a number of times already
to bound gravitational wave measurement errors. The WWB and 
ZZB on the other hand are computationally less simple, and 
we apply them here to gravitational wave parameter estimation 
for the very first time. The CRB is known as a local bound 
because it assumes that the parameters one 
seeks to estimate are deterministic, and 
provides bounds on their MSE for every possible set of 
intrinsic parameter values. The WWB and ZZB are known 
as global (Bayesian) bounds because they assume that 
the parameter set is random, of known prior distribution. 
They bound the
global MSE averaged over this prior distribution. We first
set up a model problem in order to develop intuition about 
the conditions under which global bounds are more appropriate 
than their local counterparts. Then we obtain the WWB and
ZZB for the Newtonian-form of the coalescing binary waveform, 
and compare them with published analytic CRB and numerical Monte 
Carlo results. At large signal-to-noise ratio (SNR), we find 
that the theoretical bounds are all identical and are attained 
by the Monte-Carlo results. As SNR gradually drops below $\sim 10$, 
the WWB and ZZB are both found to provide increasingly tighter 
lower bounds than the CRB. However, at these levels of moderate
SNR, there is a significant departure between all the bounds 
and the numerical Monte Carlo results. We argue that the 
WWB and ZZB are probably within a few percent of the 
theoretical minimum MSE attainable for this problem. The
implication is that the maximum likelihood method of parameter
estimation used by the Monte Carlo simulations is not the optimal 
estimator for this problem at low-to-moderate SNR. In fact, it 
is well-known that the optimal parameter estimator is the 
conditional mean estimator. This, unfortunately, is notoriously 
difficult to compute in general. We therefore advance a 
strategy for implementing this method efficiently, as a 
post-processor to 
the maximum likelihood estimator, in order to achieve improved
accuracy in parameter estimation. 
\end{abstract}
%
%
\newpage
\narrowtext \onecolumn

\section{introduction}
\label{sec;intro}

A network of very sensitive instruments is presently
being assembled across the globe with a view to directly
detecting cosmic gravitational radiation on Earth. 
Although detection is the initial goal of the network, it
must ultimately function as an astrophysical observatory. 
In this capacity it will study a range of sources of 
gravitational waves, such as neutron stars, black holes, 
and the early universe. The network will also provide
data through which to learn new and fundamental physics
about the gravitational field. Arguably, broadband 
instruments based upon the method of laser interferometry
offer the best long-term potential for making astrophysical observations. 
Several such instruments have been funded
and are being constructed at the present time. They include: LIGO, 
the US 4\,km arm-length Laser Interferometric Gravitational 
Wave Observatory \cite{ligo}; VIRGO, a French/Italian project
to construct a 3\,km arm-length interferometer \cite{virgo}; GEO600, a
UK/German effort to build a 600\,m arm-length interferometer 
\cite{geo}; TAMA, a Japanese project for the construction of 
a 300\,m arm-length interferometer \cite{tama}.  
Joint observations between these interferometers 
are scheduled for soon after the turn of the century. All of 
the instruments will continue to improve their 
sensitivity incrementally for many years thereafter.

Interferometers are intrinsically noisy instruments and
the presence of noise will mask the identity of 
all but the very strongest incident gravitational
waves. This requires a very careful and sophisticated
processing of the data, in order to extract the valuable 
information that is borne by the waves \cite{BFSBLAIR}.
Let us consider the data acquired by only a {\em single} detector 
in the network. Suppose that a detection of a gravitational
wave of some assumed known form has already been made in the 
instrument's noisy output. Our focus of attention here
is the subsequent measurement
of the one-or-more parameters that characterize the waveform. 
Specifically, we would like to compute the theoretical minimum 
mean-square errors with which the parameters can 
in principle be measured. This will clearly impact on 
the astrophysical inferences that can then be drawn as 
a result of the observation. 

The particular gravitational waveform that we will focus on in 
this paper is the {\it chirp} of 
gravitational radiation that precedes the coalescence of a
compact binary system comprising of neutron stars (NS) 
and/or black holes (BH). Coalescing binaries are the most 
promising sources of gravitational waves in the long run for 
the LIGO/VIRGO/GEO600/TAMA detectors \cite{kip300,3mins,kipsnow}. 
As radiation reaction drives 
the stars through a slow inspiral phase just prior to 
coalescence, the binary generates a very clean gravitational 
wave signal that is amenable to theoretical modeling
(see \cite{kip300,kipsnow,will94,granch,blanchet96} and references therein). 
LIGO/VIRGO anticipate observing 
the last few minutes of NS-NS inspiral, during which the 
gravitational waves oscillate through $\sim 16\,000$ cycles as 
their frequency sweeps through the visibility bands of the 
detectors. The coalescing binary event rate predictions are 
subject to gross uncertainties. However, it is not unreasonable 
that there may be a few NS-NS, NS-BH, and BH-BH mergers out to 
a distance of $\sim 200$\,Mpc  in a 
period of a year \cite{narayan,phinney,vandenh,lpp,lpp1}. 

Inspiraling binary gravitational waves are encoded with a
rich suite of physical and astrophysical information. This
ranges from tests of general relativity\cite{kipsnow,will,poisson}, 
through measurements of neutron star and black hole mass and spin
\cite{3mins,kipsnow,cf,pw,finn96}, 
to new and independent inferences about the value of the cosmological 
parameters \cite{bfsnature,mark,cherfinn,wt}. The informational 
content of binaries has driven gravitational wave theorists to 
focus much of their collective effort on waveform calculations
for inspiraling binaries. In tandem with this, there has been 
considerable study of algorithms for analyzing noisy gravitational 
wave data to extract the waveform information.  

The main goal of this paper is to re-asses the issue of 
information extraction with respect to observations of 
coalescing binaries by an interferometric gravitational 
wave detector. Our motivation is the result of a recent
confrontation between the theoretical Cramer-Rao low bound 
(CRB) on parameter estimation errors \cite{crb,vantrees} for 
coalescing binary waveforms, 
with real measurement errors based upon application of the 
{\it maximum likelihood} (ML) method of parameter estimation to 
simulated data sets \cite{sathyasim}. The ML errors were found to 
depart significantly from the CRB at moderate signal-to-noise 
ratio's (SNR) of around $\sim 8$. This implies one of the 
following: (i) the CRB is a {\em weak} lower bound at this
SNR, in which case a tighter theoretical bound would be 
desirable; (ii) the CRB is a tight bound and the ML method
is not the optimal estimator, in which case a more refined 
parameter estimation method would be desirable; (iii) the 
CRB is both weak {\em and} the ML method is not 
optimal, in which case it may be desirable to seek an 
improved theoretical lower bound and an improved 
estimator. 

In an attempt to discriminate between these options, we 
apply new theoretical bounds on measurement accuracy to 
the problem. Specifically, we investigate the Weiss-Weinstein
bound \cite{wwb} and the Ziv-Zakai bound \cite{zzb}. 
These are {\it Bayesian bounds} that
are much more versatile than the more familiar CRB. Although
a little more difficult to compute than the CRB, they can often
be considerably tighter. This has been demonstrated for
a range of parameter estimation problems in radar and sonar \cite{bell}. 

The paper is organized as follows. Sec.\,II
is an overview of parameter estimation, emphasizing
the differences between {\em local bounds} on parameter 
estimation accuracy of which the CRB is an example, 
and Bayesian bounds of which the Weiss-Weinstein
and Ziv-Zakai bounds are examples. The Weiss-Weinstein
bound in described in some detail in Sec.\,III
This is followed in Sec.\,IV by a detailed
description of the Ziv-Zakai bound. In Sec.\,V,
some of the computational issues posed by these bounds are 
presented. The bounds are applied, in Sec.\,VI, to a 
simple problem that illustrates the general superiority of 
Bayesian bounds over their local counterparts.
Then, in Sec.\,VII, the bounds are computed for a 
Newtonian coalescing binary waveform immersed in 
Gaussian random noise of spectral density characteristic
of the first-stage LIGO detectors.
We compare the bounds with actual parameter estimation errors
that were obtained recently from a Monte-Carlo experiment 
designed to test the maximum-likelihood method. 
The main results are discussed in Sec.\,VIII and some pointers
for future work are given. 

\section{summary of parameter estimation}
\label{sec;summ}

A common problem in many fields, ranging
from radar and sonar through to geophysics and astronomy, 
is to seek estimates for the set of parameters characterizing 
a waveform that is corrupted by additive Gaussian noise. 
Consider the following such observation 
\begin{equation}\label{eqn;wave}
    x(t) = s(t;\vtheta) + n(t)\,, \quad |t| < T/2\,,
\end{equation}
and assume the signal $s(t;\vtheta)$ is a known function 
of time for all values of the parameter vector $\vtheta$, 
and $n(t)$ is a zero-mean Gaussian random 
noise process. Some measurement algorithm, which we do not
need to specify in detail yet, is applied to $h(t)$ in 
order to extract $\vtheta$. The algorithm produces an 
estimate $\hat{\vtheta}$, with associated error 
$\vepsilon \equiv \vtheta-\hat{\vtheta}$. A statistical 
summary of the performance of the algorithm is contained in 
the error covariance matrix, given by 
$ {\cal R} \equiv \langle \vepsilon \vepsilon^\trp \rangle$, 
where $\langle\cdot\rangle$ denotes expectation.  
The diagonal elements of ${\cal R}$ are the mean-square
errors (MSE's) on each individual parameter, while 
the off-diagonal elements represent their cross-covariances. 
In order to benchmark the performance of any practical 
parameter estimator, one would like to know the theoretical 
{\em minimum} ${\cal R}$ for the problem at hand. 
If the signal parameters
enter nonlinearly, then a closed-form expression for 
this cannot be found, and in general it is prohibitively difficult
to compute numerically. 
A schematic picture of how the MSE will depend on SNR in a
nonlinear parameter estimation problem is given in Fig.\,1. 
In the small error or asymptotic region, characterized by 
high SNR, estimation errors are small. In the ambiguity 
region, where SNR is moderate, large errors occur. When 
SNR is very small, the observations provide little information
and the MSE is close to that obtained simply from the 
prior knowledge about the problem. In this paper we will 
be concerned with bounds that are able to characterize 
performance in the asymptotic and ambiguity regions. 
These bounds generally
fall into one of two classes: local bounds or global Bayesian
bounds. We will describe their main features in the next Section. 

\subsection{Local bounds}
\label{subsec;lb}

The formulation of local bounds is based on the premise that 
the unknown parameters one seeks to measure are deterministic 
quantities. The bounds are local in the sense that they are placed
on the MSE's for each different possible value of the intrinsic 
parameter vector. 
Local bounds have two serious limitations. First, they are 
restricted in application to estimators that are {\em unbiased}. 
In practise, biased estimation is often unavoidable. If 
the space of a parameter is finite, for example, then an unbiased 
estimator of it does not exist. Secondly, local bounds are unable 
to incorporate any prior information that one might have about the 
parameters. The Cramer-Rao bound (CRB) is a familiar example of 
a local bound and it therefore has only limited utility.  
This bound states that for any {\em unbiased} estimator 
of a parameter vector $\vtheta$, based upon noisy observations 
${\bf x}$, the error covariance matrix must be larger than or 
equal to the inverse of the Fisher information at $\vtheta$. Thus 
\begin{equation}\label{eqn;crb}
 {\cal R}_{ij} = \langle (\hat{\theta}_i-\theta_i)
                         (\hat{\theta}_j-\theta_j) \rangle
               \geq {\cal J}^{-1}_{ij}\,,
\end{equation}
where $\cal J$ is the Fisher information matrix whose 
elements are given by \cite{helstrom}
\begin{equation}\label{eqn;fisher}
  {\cal J}_{ij} =  \left\langle \frac{\partial^2}{\partial\theta_i
                                                  \partial\theta_j}
  \mbox{ln} \Lambda({\bf x};\vtheta) \right\rangle
\end{equation}
and $\Lambda({\bf x};\vtheta)$ is the likelihood ratio 
\begin{equation}\label{eqn;likrat}
\Lambda({\bf x};\vtheta) = \frac{p({\bf x}| \vtheta)}{p({\bf x}| 0)}\,.
\end{equation}
The CRB is not difficult to compute and it is widely 
invoked. In particular, it has been used almost exclusively to 
bound the measurement errors on the parameters of gravitational 
wave signals \cite{cf,pw,finncher,jk,kks,jkkt}. Moreover, it can 
be proved that the CRB is asymptotically attained by the 
maximum-likelihood (ML) method of parameter estimation 
\cite{helstrom}. As gravitational 
wave observations of coalescing binary signals will be rare, 
and the majority of detections will be made at only 
moderate SNR's, it is unlikely that the asymptotic 
conditions will be met in practise. Similar situations exist
in the fields of radar and sonar, and here alternative bounds
to the CRB have been considered. The Barankin bound, for example, is a local 
bound that can be much tighter than the CRB \cite{bb}. 
However, it is considerably more difficult to compute as it requires 
maximization over a number of free variables. Also, being a 
local bound, it still only applies to unbiased estimators and 
is unable to incorporate prior information about the parameters
if this information is available.  
                       
\subsection{Bayesian bounds}
\label{subsec;bb}

Rather than treating the unknown parameters as deterministic
quantities, Bayesian bounds treat them as random variables 
with known prior distributions. These bounds are global in 
the sense that they bound MSE's on each of the parameters, 
averaged over their prior distributions. In contrast to their
local counterparts, Bayesian bounds are not restricted in 
application to unbiased estimators. In fact, they lower
bound the performance of {\em any} estimator. Also, unlike
local bounds, they easily incorporate any prior information 
about the parameters. It is straightforward to form a Bayesian 
version of the CRB, by simply replacing the conditional 
probability density $p({\bf x} | \vtheta)$ with a joint 
probability density $p({\bf x},\vtheta)$ using Bayes theorem. 
However the Bayesian CRB is subject to a stringent regularity 
condition: it requires the prior probability density
function of the parameters, $p(\vtheta)$ to be twice differentiable. 
In the common case of parameters that have uniform priors, 
this regularity condition is obviously not met. 

Another example of a Bayesian bound is the conditional mean 
estimation bound (CMB) \cite{vantrees}. In fact this is not 
really a bound, since it can be attained by the {\it conditional 
mean estimator} (CME). This estimator achieves the minimum 
MSE and provides the benchmark against which the performance of other
estimators should be compared. The CME of a scalar parameter 
$\theta$, based upon 
noisy observations ${\bf x}$, is given by 
\begin{equation}\label{eqn;cme}
\hat{\theta}({\bf x}) = \langle \theta | {\bf x} \rangle = 
\int_{-\infty}^\infty \theta\, p(\theta | {\bf x}){\rm d}\theta\,.
\end{equation} 
In the context of gravitational wave parameter estimation, 
the CME has been referred to as the Kallianpur-Striebel 
or nonlinear filter \cite{davis}. Unfortunately, the 
CMB is prohibitively difficult to compute for all but the 
simplest of problems. It generally requires multi-dimensional 
integrations to be performed numerically over the prior 
parameter space. 

The complexity of the CMB has motivated the formulation of
two further important Bayesian bounds --- the Weiss-Weinstein
bound and the Ziv-Zakai bound. These trade off some of the
computational complexity of the CMB, and yet are only apparently
a little less tight. In a range of applications the bounds have 
demonstrated this utility \cite{bell}. We now describe each of
these bounds in some detail in the following two Sections.

\section{weiss-weinstein bound}
\label{sec;wwb}

The Bayesian form of the CRB and the CMB discussed in the 
previous Section belong to a general class of Bayesian 
bounds that Weiss and Weinstein were able to derive
from the Schwarz inequality \cite{wwb}. An outline of their 
derivation is given here. The WWB is a member of this 
general class, but it is free from the problems that 
limit the CRB and CMB. It therefore is of much more
general utility than the latter two bounds. 
A version of the WWB for the 
case of a single parameter is obtained below. A 
statement of the multiple-parameter generalization of 
the scalar bound follows.  
   
\subsection{Single parameter}
\label{subec:wwsp}

A lower bound on the error in estimating a scalar parameter 
$\theta$, based upon noisy observations ${\bf x}$, is sought. 
Let $p({\bf x},\theta)$ denote the joint probability density
of ${\bf x}$ and $\theta$. Weiss and Weinstein introduced a 
function $\psi({\bf x},\theta)$ such that
\begin{equation}\label{eqn;orthg}
 \int_{-\infty}^\infty {\rm d}\theta\, \psi({\bf x},\theta)
                       p({\bf x},\theta)=0 \quad \forall {\bf x}\,.
\end{equation}
Since, for any real-valued measurable function $g({\bf x})$,
\begin{equation}
 \langle g({\bf x})\psi({\bf x},\theta) \rangle = 
 \int_{-\infty}^\infty {\rm d}{\bf x}\, g({\bf x}) 
 \int_{-\infty}^\infty {\rm d}\theta\,
\psi({\bf x},\theta)p({\bf x},\theta) = 0\,,
\end{equation}
the condition in (\ref{eqn;orthg}) implies that $\psi({\bf x},\theta)$ 
is orthogonal to any transformation of the data ${\bf x}$. Subtracting 
$\langle \theta\psi({\bf x},\theta) \rangle$ from both sides above
and then applying Schwarz's inequality to the left side, results in
the following 
\begin{equation}\label{eqn;coveq}
 \langle [\theta - g({\bf x})]^2 \rangle \geq
 \frac{ \langle \theta\psi({\bf x},\theta) \rangle^2 }
      { \langle \psi^2({\bf x},\theta) \rangle }\,.
\end{equation}
As this inequality is valid for any $g({\bf x})$, it
sets a lower bound on the mean-square error in estimating $\theta$
from observation of ${\bf x}$. To underline this point we will 
replace $g({\bf x})$ in the following by $\hat{\theta}({\bf x})$. 
Note that the lower bound set by the right side of (\ref{eqn;coveq}) is 
independent of the estimator (i.e. is absent of $g({\bf x})$).
Recall that local bounds, such as 
the CRB, do not share this property of Bayesian bounds: they
apply only to estimators that are unbiased. 

It is simple to see that the Bayesian form of the CRB and the CMB 
are special cases of the general bound in (\ref{eqn;coveq}). 
Consider choosing the function $\psi({\bf x},\theta)$ as follows: 
\begin{equation}\label{eqn;psicrb}
  \psi({\bf x},\theta) = \frac{\partial\mbox{ln}\;p({\bf x},\theta)}
                         {\partial\theta}\,.
\end{equation}
This choice satisfies the orthogonality condition (\ref{eqn;orthg})
and generates the Bayesian CRB. Similarly, the selection
\begin{equation}\label{eqn;psicmeb}
 \psi({\bf x},\theta) = \theta - \langle \theta | {\bf x} \rangle
\end{equation}
leads to the CMB. 

In their quest for a less restrictive bound than the CRB 
and the CMB, Weiss and Weinstein  were led to consider a different 
choice for $\psi({\bf x},\theta)$. They proposed the following 
\begin{equation}\label{psiwwb}
 \psi({\bf x},\theta) = L^r({\bf x};\theta+\delta,\theta) - 
                        L^{1-r}({\bf x};\theta-\delta,\theta)\,,
\end{equation}
where $r$ and $\delta$ are arbitrary real-valued scalars and 
$L({\bf x};\theta_1,\theta_2)$ is the {\it likelihood ratio}
\begin{equation}\label{eqn;lr}
 L({\bf x};\theta_1,\theta_2) = \frac{p({\bf x},\theta_1)}{p({\bf x},\theta_2)}\,.
\end{equation}
This choice for $\psi({\bf x},\theta)$ satisfies the orthogonality 
condition (\ref{eqn;orthg}) for all combinations of $\delta$ and 
$0<r<1$. Substitution into Eq. (\ref{eqn;coveq}) generates the 
WWB on the mean-square
error, $\epsilon^2$, in the estimation of $\theta$:
\begin{equation}\label{eqn;wwb}
 \epsilon^2 \geq 
 \frac { \delta^2 \exp[2\eta(r,\delta)] }
       { \exp[\eta(2r,\delta)] + \exp[\eta(2-2r,-\delta)]
                             - 2\exp[\eta(r,2\delta)] }\,,
\end{equation}
where
\begin{eqnarray}\label{eqn;eta}
 \eta(r,\delta) &=& \mbox{ln}\;\langle L^r({\bf x};\theta+\delta,
                \theta) \rangle \nonumber \\ 
                &=& \mbox{ln}\;\int_\Theta p^r(\theta+\delta)
                                        p^{1-r}(\theta)
 \left\{ \int p^r({\bf x} | \theta+\delta) 
 p^{1-r}({\bf x} | \theta){\rm d}{\bf x}
 \right\} {\rm d}\theta   \nonumber     \\
                &=& \mbox{ln}\int_\Theta p^r(\theta+\delta)
 p^{1-r}(\theta)\exp[\mu(r;\theta+\delta,\theta)]{\rm d}\theta\,.
 \end{eqnarray}
Several comments are pertinent here. First, note that the integration
with respect to $\theta$ is performed over the region 
$\Theta=\{\theta:p(\theta)>0\}$ in order to avoid singularities. 
Second, the bound reduces to the Bayesian version of the CRB 
for $\delta \rightarrow 0$. Third, the term $\mu(r;\theta+\delta,\theta)$
is a familiar one in information and communications theory: it is 
known as the semi-invariant moment generating function and used
to bound the probability of error in binary hypothesis testing
problems \cite{vantrees}. We shall meet it again in our discussion of the 
Ziv-Zakai bound in the next Section.  

In order to remove one of the degrees of freedom, the
WWB is usually computed for $r=1/2$. It then reduces to
\begin{equation}\label{eqn;wwb1}
 \epsilon^2 \geq \frac {\delta^2 \exp[2\eta(1/2,\delta)]}
                       {2\{1-\exp[\eta(1/2,2\delta)]\}}\,.
\end{equation}
The variable $\delta$ that enters the bound is usually 
referred to as a test point. The optimal value for 
$\delta$ is the one 
that generates the maximum bound. This value may be
other than $\delta \rightarrow 0$, for which the WWB reduces
to the CRB as we remarked earlier.   

Weiss and Weinstein have shown how to generalize the single test point
bound (\ref{eqn;wwb1}) to incorporate multiple test points. 
Consider a vector of $N$ test points
$\vdelta \equiv (\delta_1,\delta_2, \ldots, \delta_N)$.
The corresponding multiple-test point WWB is
\begin{equation}\label{eqn;wwb3}
 \epsilon^2 \geq {\bf u} {\cal Q}^{-1} {\bf u}^\trp\,,
\end{equation}
where the elements of the vector {\bf u} are
\begin{equation}\label{eqn;uvec}
u_i = \delta_i \,,
\end{equation}
and the elements of the matrix ${\cal Q}$ are 
\begin{equation}\label{eqn;qmat}
{\cal Q}_{ij} = 2\frac{ \exp[\eta(1/2,\delta_i-\delta_j)]-
                    \exp[\eta(1/2,\delta_i+\delta_j)]}
                    {\exp [2\eta(1/2,\delta_i)]}\,.
\end{equation}
In order to evaluate Eq. (\ref{eqn;wwb3}), a matrix of
dimension equal to the number of test points has to be
inverted numerically. This imposes a practical restriction
on the number of test
points. However, as we shall see later, the WWB 
fortunately appears to converge quickly with increasing $N$. 

\subsection{Multiple parameters}
\label{subsec:wwmp}

Since the coalescing binary waveform is characterized by
more than one parameter, we shall require a
multiple parameter version of the WWB. Consider a vector of 
$M$ parameters, $\vtheta \equiv (\theta_i, \ldots, \theta_M)$.
The WWB on the error covariance matrix $\cal R$ is obtained in 
a similar fashion to the single parameter bound. The result is,   
\begin{equation}\label{eqn;wwb4}
 {\cal R} \geq {\cal H} {\cal G}^{-1} {\cal H}^\trp\,.
\end{equation}
The elements of the matrix ${\cal H}$ are the 
$M \times N$ test points in the multi-dimensional 
parameter space. The $N \times N$ matrix 
${\cal G}$ has elements given by
\begin{equation}\label{eqn;gmat}
 {\cal G}_{ij} = 2 \frac { \exp[\eta(1/2,\vdelta_i-\vdelta_j)] - 
                      \exp[\eta(1/2,\vdelta_i+\vdelta_j)] }
                    { \exp[\eta(1/2,\vdelta_i)] \,
                     \exp[\eta(1/2,\vdelta_j)] }\,,
\end{equation}
where $\vdelta_i$ is the $i$'th test point in the 
parameter space and
\begin{equation}\label{eqn;eta1}
 \eta(1/2,\vdelta_i\vdelta_j) = \mbox{ln}\;\int_\Theta
 \sqrt{ p(\vtheta+\vdelta_i)p(\vtheta+\vdelta_j) }
 \left\{ \sqrt{ p({\bf x}|\vtheta+\vdelta_i)
           p({\bf x}|\vtheta+\vdelta_j) }{\rm d}{\bf x} \right\} 
           {\rm d}\vtheta\,.
\end{equation}
Again, several comments are pertinent. 
First, integration with respect to 
$\vtheta$ is over the region $\Theta=\{\vtheta : p(\vtheta)>0\}$.
Second, for a non-singular bound there must be at least 
$M$ linearly independent test points. Finally, (\ref{eqn;wwb4})
reduces to the Bayesian CRB upon setting ${\cal H}=\delta{\cal I}$, 
where $\cal I$ is the identity matrix and the scalar 
$\delta \rightarrow 0$. We will turn to the practical issues 
involved in the computation of the scalar and vector parameter
versions of the WWB in Sec.\,V.

\section{Ziv-Zakai Bound}
\label{sec;zzb}

The theoretical foundation of the Ziv-Zakai bound (ZZB) 
is somewhat different to the WWB \cite{zzb}: there do not appear to be 
any formal theoretical links. In common
with the WWB however, the ZZB places a fundamental lower bound
on the performance of {\em any} parameter estimator. We
present a simple derivation here, for the case of              
a single parameter with a uniform prior distribution.
The multiple-parameter extension of the bound for 
arbitrary priors is also presented.

\subsection{Single parameter}
\label{subsec;zzsp}

As a concrete example, suppose that an 
estimate of the difference between the arrival time
of a gravitational wave at two separated detectors is required. 
Let us denote this parameter by $\theta$.  Now
ask what is the probability of making a correct 
decision between two possible values, $\phi$ and 
$\phi+\Delta$, of this parameter. The likelihood
ratio test (LRT) is the optimal decision scheme that 
produces the minimum probability of error. Instead of
the LRT, consider a simpler suboptimal decision scheme 
in which a decision is made in favour of 
the ``nearest-neighbour'' to some arbitrary 
estimate, $\hat{\theta}$, of $\theta$. Thus, 
\begin{eqnarray}\label{eqn;subopt}
 \mbox{Decide} \; H_0: \theta=\phi &\quad \mbox{if} \quad & 
              \hat{\theta} \leq \phi+\frac{\Delta}{2}\,, \nonumber \\
 \mbox{Decide} \; H_1: \theta=\phi+\Delta &\quad \mbox{if} \quad &
              \hat{\theta} > \phi+\frac{\Delta}{2}\,.
\end{eqnarray}
If the two hypothesized delays are equally likely to 
occur, which is physically most reasonable, then the 
suboptimal decision scheme has a probability of 
error given by
\begin{eqnarray}\label{eqn;perr}
 P(\phi,\phi+\Delta) = &\frac{1}{2}&P\left(\hat{\theta}>\phi+
 \frac{\Delta}{2}\Biggl|\Biggr.\,\theta=\phi\right) \nonumber \\
                     + &\frac{1}{2}&P\left(\hat{\theta}\leq \phi+
 \frac{\Delta}{2}\Biggl|\Biggr.\,\theta=\phi+\Delta\right)\,.
\end{eqnarray}
Clearly if $P_{\rm min}(\phi,\phi+\Delta)$ is the minimum 
probability of error, associated with the LRT, then  
\begin{eqnarray}\label{eqn;pineq}
 P_{\rm min}(\phi,\phi+\Delta) \leq &\frac{1}{2} &P\left(\epsilon>
 \frac{\Delta}{2}\biggl|\biggr.\,\phi\right) \nonumber\\
                                   +&\frac{1}{2} &P\left(\epsilon \leq
 -\frac{\Delta}{2}\biggl|\biggr.\,\phi+\Delta\right)\,,
\end{eqnarray}
where $\epsilon=\hat{\theta}-\theta$ denotes the estimation error.  
Now, suppose that $\theta$ is uniformly distributed on $[-T,T]$. 
In this specific example, $T$ would represents the gravitational wave
travel time between the two detectors. The inequality
(\ref{eqn;pineq}) holds good for any $\phi$ and $\Delta$, in  
particular combinations of $\phi$ and $\Delta$ such that 
$\phi,\phi+\Delta \in [-T,T]$, or
\begin{equation}\label{eqn;prior}
 -T \leq \phi \leq T-\Delta, \quad \quad 0 < \Delta < 2T\,.
\end{equation}
Integrating Eq. (\ref{eqn;pineq}) with respect to $\phi$ over
$[-T,T-\Delta]$ gives
\begin{equation}\label{eqn;step1}
 \int_{-T}^{T-\Delta} P_{\rm min}(\phi,\phi+\Delta) {\rm d}\phi\leq 
 \frac{1}{2}\int_{-T}^T 
P\left((|\epsilon|\geq \frac{\Delta}{2} \biggl|\biggr. \phi\right)\,
{\rm d}\phi\,,
\end{equation}
which can equivalently be expressed as
\begin{equation}\label{eqn;step2}
 \int_{-T}^{T-\Delta} P_{\rm min}(\phi,\phi+\Delta){\rm d}\phi \leq
 T\cdot H\left(\frac{\Delta}{2}\right)\,,
\end{equation}
where
\begin{equation}\label{eqn;step2a}
 H(\Delta) \equiv 
 \frac{1}{2T} \int_{-T}^T P(|\epsilon|\geq \Delta|\,\phi)\, {\rm d}\phi\,.
\end{equation}
Note that (\ref{eqn;step2}) is only useful for $\Delta \leq 2T$, 
since for $\Delta > 2T$ the integral is negative and therefore
zero is a better bound. The next step is to multiply both 
sides of (\ref{eqn;step2}) by $\Delta/T$ and integrate with 
respect to $\Delta$ over $[0,2T]$. Noting that  
\begin{equation}\label{eqn;step3}
 \epsilon^2 = -\int_0^{2T} \Delta^2 {\rm d}\{H(\Delta)\}
\end{equation}
is the mean-square error in the estimation of 
$\theta$ when the latter has a uniform prior distribution in 
$[-T,T]$, the integration yields: 
\begin{equation}\label{eqn;step4}
 \epsilon^2 \geq \frac{1}{2T} \int_0^{2T} \Delta\, {\rm d}\Delta
 \int_{-T}^{T-\Delta} P_{{\rm min}}(\phi,\phi+\Delta){\rm d}\Delta\,,
\end{equation}
which is the ZZB in its simplest incarnation. Bellini and 
Tartara \cite{valleyfn} have remarked that $H(\Delta)$ is a 
non-increasing function of $\Delta$ and suggested that 
the bound might be tightened by applying a `valley-filling' 
function to the left-side of (\ref{eqn;step2}). Denoting
this function by $V[\cdot]$, the Bellini-Tartara version 
of the ZZB is
\begin{equation}
 \epsilon^2 \geq \frac{1}{2T} \int_0^{2T} \Delta \cdot
 V\left[ \int_{-T}^{T-\Delta} 
 P_{{\rm min}}(\phi,\phi+\Delta){\rm d}\phi\right]{\rm d}\Delta\,.
\end{equation}
The bound generalizes in a straightforward manner
for an arbitrary prior, $p(\theta)$, to give
\begin{equation}
 \epsilon^2 \geq \int_0^\infty \frac{\Delta}{2} \cdot V\left\{
 \int_{-\infty}^\infty \left[p(\phi)+p(\phi+\Delta)\right]\cdot
 P_{\rm {min}}(\phi,\phi+\Delta){\rm d}\phi\right\}{\rm d}\Delta\,.
\end{equation}
Although there is generally no closed form expression for 
$P_{{\rm min}}(\phi,\phi+\Delta)$, tight lower
bounds exist \cite{vantrees}.   

\subsection{Multiple parameters}
\label{subsec:zzmb}

The ZZB has only recently been extended to vector random 
parameters with arbitrary prior distributions \cite{bell}. Consider
an $M$-dimensional vector random variable, $\vtheta$, with 
prior pdf $p(\vtheta)$. As before, let $\hat{\vtheta}$ be
an estimate of $\vtheta$ produced by any estimator,  
$\vepsilon$ the estimation error, and 
${\cal R}=\langle \vepsilon\vepsilon^\trp \rangle$ 
the error covariance matrix. 
Then the following lower bound 
on ${\bf a}^\trp {\cal R} {\bf a}$ for any $M$-dimensional
vector ${\bf a}$ has been obtained: 
\begin{equation}\label{eqn;vzzb}
 {\bf a}^\trp {\cal R} {\bf a} \geq \\
 \int_0^\infty \frac{\Delta}{2} \cdot V'\,{\rm d}\Delta\,,
\end{equation}
where the valley filling function $V'$ is now defined as
\be
V'\equiv V\left\{
 \mbox{max}\,
 \int( p(\vphi)+p(\vphi+\vdelta) ) \cdot 
 P_{\rm min}(\vphi,\vphi+\vdelta){\rm d}\vphi \right\}\,,
\end{equation}
with the maximum referred to $\vdelta$. 
The bound is generated, as for the single parameter case, via
an inequality between the probability of error in a suboptimal
decision rule and the minimum probability of error associated
with an LRT. However, one has now to decide between one of two possible
values $\vphi$ or $\vphi+\vdelta$ for the parameter vector
under investigation. 
The suboptimal decision rule is then:
\begin{eqnarray}\label{eqn;subopt1}
 \mbox{Decide} \, H_0: \vtheta=\vphi & \quad \mbox{if}\quad & 
              {\bf a}^\trp\hat{\vtheta}>{\bf a}^\trp\vphi+\frac{\Delta}{2}\,,
\nonumber \\
 \mbox{Decide} \, H_1: \vtheta=\vphi+\vdelta & \quad \mbox{if} \quad &
              {\bf a}^\trp\hat{\vtheta}\leq {\bf a}^\trp\vphi+\frac{\Delta}{2}
\,. 
\end{eqnarray}
The hyperplane
\begin{equation}
 {\bf a}^\trp{\vtheta} = {\bf a}^\trp\vphi + \frac{\Delta}{2}\,,
\end{equation}
separating the two decision regions, passes through the midpoint
of the line connecting $\vphi$ and $\vphi+\vdelta$ and is 
perpendicular to the ${\bf a}$ axis. A decision is made in favour
of the hypothesis that is on the same side of the separating
hyperplane as the estimate $\hat{\vtheta}$. The tightest bound
in (\ref{eqn;vzzb}) is achieved by maximization over the 
vector $\vdelta$, subject to the constraint 
${\bf a}^\trp\vdelta=\Delta$. This constraint does not 
determine the vector $\vdelta$ uniquely. In order to satisfy it, 
$\vdelta$ must be composed of a fixed component along the 
${\bf a}$ axis, and an arbitrary component orthogonal to 
${\bf a}$. That is
\begin{equation}
 \vdelta = \frac{\Delta}{\parallel {\bf a} \parallel^2} {\bf a}+{\bf b}\,,
\end{equation}
where
\begin{equation}
 {\bf a}^\trp {\bf b} = 0\,, 
\end{equation}
and there are $M-1$ degrees of freedom in choosing $\vdelta$
via the vector ${\bf b}$. Simply setting ${\bf b}=0$ results
in hypotheses that are separated by the smallest Euclidean distance.
However, this does not necessarily guarantee the largest probability of 
error. A maximization over $\vdelta$ can improve the bound.  
                   
\section{Computational Issues}
\label{sec;comp}

In this Section the CRB, WWB, and ZZB are reduced to   
forms that are appropriate for calculating error  
bounds on the parameters of a signal that is immersed 
in a background of {\em Gaussian} random noise. 

The signal waveform $s(t,\vtheta)$ is parameterized
by a vector, $\vtheta$, for which an estimate is sought. 
Noisy measurements of the signal are obtained as follows:
\begin{equation}
 x(t) = s(t,\vtheta) + n(t)\,, \quad  
 -\frac{T}{2} \leq t \leq \frac{T}{2}\,,
\end{equation}
where $n(t)$ is the noise, assumed Gaussian with 
a known spectral density denoted by $S_n(f)$. 
 
\subsection{The Cramer-Rao Bound}
The CRB was defined in (\ref{eqn;crb}) to be the inverse of the 
Fisher information at $\vtheta$.  
The latter is a matrix 
of second derivatives of the likelihood ratio of an 
observation with respect to $\vtheta$. 
In the case of stationary Gaussian noise, the matrix 
elements reduce to
\begin{equation}\label{eqn;fimg}
 {\cal J}_{ij} =  \left( \frac{\partial s}{\partial\theta_i}\biggl|\biggr.
                    \frac{\partial s}{\partial\theta_j} \right)\,,
\end{equation}
where $(s_1 | s_2)$ denotes the inner product between two 
signals, $s_1(t)$ and $s_2(t)$. In terms of the signal's
Fourier transforms, $\tilde{s}_1(f)$ and $\tilde{s}_2(f)$, 
and the spectral density of the noise, $S_n(f)$, the 
inner product can be expressed as
\begin{equation}\label{eqn;innpr}
 (s_1 | s_2) = 2 \int_0^\infty \frac
 {\tilde{s}_1^*(f)\tilde{s}_2(f)+\tilde{s}_1(f)\tilde{s}_2^*(f)}
 {S_n(f)} {\rm d}f\,.
\end{equation}
The integral is a measure of the degree of `overlap' between the 
two signals, and in radar applications it is often termed 
the {\it ambiguity function}. Thus, (\ref{eqn;fimg}) can be interpreted 
as the local curvature of the signal ambiguity
function around its maximum. Numerical integration is 
generally required to compute the elements of the 
Fisher information matrix. It is
then straightforward to perform the inversion
and obtain the CRB. The diagonal 
elements of the inverted Fisher matrix are the 
Cramer-Rao bounds on the variances of each of the signal's
parameters. 

\subsection{Weiss-Weinstein Bound}
\label{subsec;compww}
 
The calculation of the WWB relies upon evaluating 
the semi-invariant moment generating function 
$\eta(1/2;\vdelta_i,\vdelta_j)$ in (\ref{eqn;eta1}).
It is not difficult to show (see Appendix in \cite{wwb} 
for details) that, for stationary and Gaussian noise, 
this function can be reduced to
\begin{equation}\label{eqn;twoterm}
 \eta(1/2;\vdelta_i,\vdelta_j) = \mbox{ln}\;C(\vdelta_i,\vdelta_j)
                                +\mu(1/2;\vdelta_i-\vdelta_j)\,,
\end{equation}
where
\begin{equation}
 C(\vdelta_i,\vdelta_j) = \int_\Theta 
 \sqrt{ p(\vtheta+\vdelta_i)p(\vtheta+\vdelta_j) }{\rm d}\vtheta\,,
\end{equation}
and the region of integration is $\Theta=\{\vtheta:p(\vtheta)>0\}$.
In Eq. (\ref{eqn;twoterm}),
the first term embodies the prior information about the 
parameters. Consider the single test-point version of the WWB
for a single parameter having a uniform prior on the interval 
$[-D/2,D/2]$. The integral is simple to evaluate and it yields  
\begin{equation}
 C(\delta) = {\rm ln}\, \left( 1 - \frac{|\delta|}{D} \right)\,.
\end{equation}
After some algebra, the second term in (\ref{eqn;twoterm}) reduces to 
\begin{equation}\label{eqn;mu}
 \mu(1/2;\vdelta_i-\vdelta_j) = -\frac{1}{4}\rho^2[1-
                                 \gamma(\vdelta_i-\vdelta_j)]\,,
\end{equation}
where $\rho^2 \equiv (s|s)$ is the squared amplitude (energy)
signal-to-noise ratio, and 
\begin{equation}
 \gamma(\vdelta_i-\vdelta_j) = \frac{\left(s[\vtheta+\vdelta_i-\vdelta_j]\;
                                     \big|\;s[\vtheta]\right) }
                              {\left(s[\vtheta] \;\big|\;s[\vtheta]\right)}
\end{equation}
is the normalized signal ambiguity function. It is often the 
case in practise that the latter function is independent of 
$\vtheta$, and then it's calculation is greatly simplified. 
This will be the case for the examples that are presented later.
However, numerical integrations are still generally required to 
compute the signal ambiguity function.  
Moreover, these integrations have to be performed for 
every set of test point locations in the parameter space.
As the WWB also requires
inversion of a matrix having dimension equal to the 
number of test points, it is clearly desirable to 
keep the number of test points down to a minimum. An indicator 
of the number of test points that are required for a 
given problem, and their optimal locations, is the shape of
the signal ambiguity function. As we shall see later, for
the coalescing binary waveform this function has a very 
well-defined shape. 

\subsection{The Ziv-Zakai Bound} 

The main term on which the evaluation of the ZZB in (\ref{eqn;vzzb})
hinges is $P_{\rm min}$, the minimum probability of error in a 
binary detection problem. An exact expression for $P_{\rm min}$
exists for the decision problem of discriminating between
two {\em equally likely} signal 
vectors, ${\bf s}_1$ and ${\bf s}_2$, in a background of Gaussian 
noise of covariance ${\bf K}$. The minimum probability of error 
is then given simply by 
\begin{equation}\label{eqn;pmin}
 P_{\rm min} = \Phi\left(\frac{d}{2}\right)\,,
\end{equation}
where $d$ is the normalized distance between the signals
\begin{equation}\label{eqn;dist}
 d = \sqrt{ ({\bf s}_2-{\bf s}_1)^\trp {\bf K}^{-1} 
            ({\bf s}_2-{\bf s}_1) }\,,
\end{equation}
and
\begin{equation}\label{eqn;errfn}
\Phi(z) = \int_z^\infty \frac{1}{\sqrt{2\pi}}\,e^{-t^2/2}
{\rm d}t\,.
\end{equation}  
If the inner products under the square root sign  
in (\ref{eqn;dist}) are evaluated, one finds that
\begin{equation} \label{eqn;dist1}
 d = \sqrt{-\mu(1/2;\vdelta_i-\vdelta_j)}\,,
\end{equation}
where $\mu$ is given by Eq. (\ref{eqn;mu}).
Therefore the calculation of the ZZB, like the WWB, is 
crucially dependent on the shape of the signal 
ambiguity function. In the case of the WWB this  
dictates the number of test points and their locations 
in order to achieve a tight bound. In terms of 
the ZZB, the shape of the signal ambiguity function 
defines a path of integration in (\ref{eqn;vzzb}),
subject to the constraint under which the integration is evaluated.

\section{An Illustrative Example}
\label{sec;ex}

In this Section we apply the CRB and one of the 
Bayesian bounds (WWB) to a  
simple scalar parameter estimation problem. Our 
intention is to investigate the conditions under
which Bayesian bounds are tighter than local bounds. 

A common feature of all the bounds that we have
presented is that they depend on the shape of the  
signal ambiguity function, $\gamma$, rather than the signal shape,   
when the background of noise is Gaussian. Often the
shape of the signal ambiguity function is the same
for all underlying values of the signal's parameters, greatly
simplifying the calculation of the bounds. 

The CRB only probes the shape of the signal ambiguity
function around it's maximum. Structure in the 
ambiguity function away from the maximum could be 
enhance by noise and masquerade as a false peak. 
This would confound a maximum likelihood parameter 
estimator, and may lead to numerical parameter estimation
errors that depart significantly from the theoretical CRB. 

While the CRB is `blind' to the presence of sidebands
in the signal ambiguity function, the WWB and ZZB are
able to capture this structure. In the case of the WWB
this is achieved through the test points. As well as  
probing around the main lobe of the ambiguity function, 
test points may also be placed around the secondary maxima. 
Similarly, the ZZB is generally tighter than the CRB if
a path of integration is selected to traverse all of
the predominant lobes in $\gamma$. 

The difference between the CRB and the WWB is best 
illustrated through an example. We consider two 
signals, $s_1$ and $s_2$, characterized by a 
scalar parameter, $\theta$, that enters the signals
nonlinearly. The signal waveforms need not concern  
us here, only their ambiguity functions
$\gamma_i(\theta,\Delta\theta)$, where $i = 1,2$.
We have chosen signals whose $\gamma$'s
are independent of $\theta$ and depend only on 
displacements in $\theta$, {\em i.e.} $\Delta\theta$. 
These ambiguity functions are displayed in Fig.\ref{fig;toy}.  
The signal $s_1$ was designed so that  
$\gamma_1$ has only a single broad maximum. 
The other signal, $s_2$, has
$\gamma_2$ comprising of a number of significant
secondary lobes. Note also that $\gamma_1$ is
actually the ``envelope" of $\gamma_2$. 

We assume that $\theta$ has a uniform prior. However, 
the region of support of the prior is set much larger than the 
anticipated parameter estimation errors, so that the prior 
does not actually impact upon the estimation accuracy for this problem.   

For $s_1$, the WWB was computed by placing $N$ test points
uniformly along the lobe of the signal ambiguity function.  
It was found that
the resulting bound was not very sensitive to where
the test points were placed along the lobe for this signal.  
A variable number of test points (up to 20) were used
to study the convergence of the bound. This was attained 
for only $4$ test points. 
For $s_2$, the test
points were placed around the main lobe of the ambiguity 
function and also around the principal secondary maxima. 
In fact only the first three secondary lobes needed
to be covered: again the WWB exhibited rapid convergence.
 
The CRB was also obtained for $s_1$ and $s_2$ over an 
identical range of SNR. This was obtained by  
inverting the Fisher matrix (\ref{eqn;fimg}) for this specific problem. 
However, the 
CRB can also be computed in terms of the WWB formalism 
for a single test point, $\delta$, allowing 
$\delta \rightarrow 0$. It should be remarked that in  
the multiple test-point formulation of the WWB, one 
test point is always forced to have $\delta\rightarrow 0$. 
This ensures that the WWB reduces to the CRB at large SNR. 

The bounds, 
$\epsilon_{\theta}^{(i)}$ ($i=1,2$), that we calculated
are displayed 
in the top panel of Fig. \ref{fig;toy}. For $s_1$, the WWB 
and the CRB are in good agreement down to SNR $\simeq 11$. 
At smaller values of SNR, the WWB is a tighter bound
than the CRB but not significantly so. This result
is not too surprising: $\gamma_1$ has only a single
broad maximum and the CRB probes the curvature of this lobe.   
Similarly the WWB probes the structure in the
ambiguity lobe, although it is able to probe  
further away from the maximum with respect to the CRB. 
As the lobe is broad, the WWB is generally a little 
tighter than the CRB. 
The results for $s_2$ are significantly different. 
Here, the WWB departs from the CRB at a much
higher value of SNR, around $20$. This is because
the CRB is blind to the secondary maxima in $\gamma_2$
that the WWB is able to capture through a judicious 
choice of test points. 
The discrepancy between the CRB and the 
WWB for this example is striking: a factor of $\sim 5$
at ${\rm SNR} = 10$ and more than an order of magnitude at ${\rm SNR} = 5$.
The WWB falls significantly as SNR increases, while the CRB 
remains fairly
constant, due to the sharp maximum in  $\gamma_2$ .

It is also interesting to compare the behaviour of 
$\epsilon_{\theta}^{(1)}$
and $\epsilon_{\theta}^{(2)}$.
At high SNR the accuracy in the determination
of $\theta$ for $s_2$ is better than for $s_1$.  
This is clearly due to the sharp maximum in $\gamma_2$. At low SNR, we expect 
$\epsilon_{\theta}^{(1)}\sim \epsilon_{\theta}^{(2)}$ because
the ambiguity functions have roughly the same 
``global" profile. This intuition is borne out in the results of the 
Bayesian analysis (WWB), but not for the local analysis (CRB)
where $\epsilon_{\theta}^{(2)} < \epsilon_{\theta}^{(1)}$ 
at all SNR. 

\section{Application to coalescing binary parameter estimation}
\label{sec;gw}

We are now in a position to investigate whether Bayesian bounds 
provide a tighter constraint than the CRB on the error covariance matrix
for the parameters of a gravitational 
wave signal generated during the inspiral phase of a compact object binary. 
In the following we use units where $G=c=1$. This 
implies a conversion factor: $1{\rm M}_\odot = 4.926\times 10^{-6}$\,s. 

\subsection{signal and noise model}
The coalescing binary inspiral waveform can be cast in the following 
generic form
\begin{equation}
 h(t) = A[\pi f(t)]^{2/3}\cos[\phi(t)]\,,
\label{ht}
\end{equation}
where $f(t)$ is the instantaneous gravitational wave frequency,
$\phi(t)$ the instantaneous phase, and $A$ the amplitude.
We will consider the phasing of the wave, as well as the
amplitude evolution, only up to Newtonian order. In this approximation
the factor $A$ is a constant, complicated, function of the binary's
distance, location in the sky,
{\it chirp mass} ${\cal M} \equiv m_1^{3/5} m_2^{3/5}/(m_1 + m_2)^{1/5}$ 
(where
$m_1$ and $m_2$ are the masses of the compact objects)
and the detector's antenna
pattern \cite{kip300}. It's precise functional form 
need not concern us further.  Frequency and phase
read
\begin{equation} 
 f(t) = f_a\left[ 1 - \frac{t-t_a}{\tau} \right]^{8/3}
\label{ft}
\end{equation}
and
\begin{equation}
 \phi(t) = \frac{16\pi f_a\tau}{5}\left\{
  1 - \left[\frac{f(t)}{f_a} \right]^{-5/3} \right\} + \Phi_a\,,
\label{phit}
\end{equation}
where the constant $\tau$, sometimes referred
to as the chirp time, can be cast in terms 
of the chirp mass of the binary as 
\begin{equation}
 \tau = \frac{5}{256} {\cal M}^{-5/3}(\pi f_a)^{-8/3}\,.
\end{equation}
The constants $f_a$ and $\Phi_a$ are respectively the 
frequency and phase of the signal at the arbitrary time $t=t_a$. The
waveform is characterized in terms of 3 parameters: 
$t_a, \Phi_a$, and $\tau$. The amplitude parameter $A$ enters
the waveform linearly and it will be
incorporated later into our definition
of signal-to-noise ratio (see Eq. (\ref{snr})). Of course, 
the parameterization of the signal
is not unique and one can express $h(t)$ as a function, 
for example, of the time to coalescence and the phase of the 
wave at this time. In fact, the latter parameterization may have 
some advantages.  However, our main goal here is to assess
the {\em relative} difference between the CRB and 
Bayesian bounds and so this detail of the parameterization is not 
crucial: we require only consistency in the choice of 
parameterization of the signal in order to compare the bounds. 
In particular the set of parameters that we are assuming here 
may not correspond to physical ones, and this would be
the case if we extended this setting to post-Newtonian waveforms
\cite{sathyasim}. However, there 
is one crucial feature of the wave parameterization 
that we adopt here: it produces a signal ambiguity function that is 
independent of the intrinsic values of the parameters
and depends only upon their displacements. 
This fact is 
more transparent if we examine the signal's Fourier transform
${\tilde h}(f)$, which in the stationary phase approximation \cite{cf}
reads:
\begin{equation}
 {\tilde h}(f) = {\cal N}f^{-7/6}\,\exp\left[\Psi(f)\right]\,,
\label{eqn;hf}
\end{equation}
where
\begin{equation}
 {\cal N} = A\pi^{2/3} \left( \frac{2\tau}{3} \right)^{1/2}f_a^{4/3}
\label{nf}
\end{equation}
is a normalization constant, and
\begin{equation}
\Psi(f) = i\sum_{\nu=1}^3 \psi_\nu(f)\lambda^\nu - i\frac{\pi}{4}\,;
\label{psif}
\end{equation}
$\lambda^\nu$ represents the parameter vector 
\begin{equation}
 \lambda^\nu\equiv \left(t_a,\Phi_a,\tau\right)\,,
\label{parchirp}
\end{equation}
and
\begin{eqnarray}
 \psi_1 &=& 2\pi f\,,\nonumber \\
 \psi_2 &=& -1\,, \nonumber \\
 \psi_3 &=& 2\pi f - \frac{16\pi f_a}{5} + \frac{6\pi f_a}{5}
                     \left( \frac{f}{f_a} \right)^{-5/3}\,.
\label{eqn;psichirp}
\end{eqnarray}
Notice that the signal's parameters enter linearly into 
the phase (\ref{psif}) of its Fourier transform. 
The normalized ambiguity function for the signal (\ref{eqn;hf}) 
is given by
\begin{equation}
\gamma(\Delta\lambda^{\nu}) = J^{-1}\int
\frac{f^{-7/3}}{S_n(f)}\cos\left\{{\cal R}e
\left[\sum_{\nu=1}^3 \psi_\nu(f)\Delta\lambda^\nu\right]\right\}\,{\rm d}f\,,
\end{equation}
where
\be
J = \int
\frac{f^{-7/3}}{S_n(f)}\,{\rm d}f\,
\ee
and the integral is defined over the frequency interval, 
within the instrument's sensitivity band, spanned by the
signal. The optimal signal-to-noise ratio reads:
\be
\rho^2 \equiv (h|h) =  4{\cal N}^2 J
\label{snr}
\ee
and, therefore, incorporates the amplitude 
parameter $A$ via the definition of ${\cal N}$, cfr. Eq. (\ref{nf}).

Our model for the noisy spectral density, $S_n(f)$, is  
intended to be representative of the performance
of the first stage
LIGO detectors. An analytic fitting formula for this has been
presented in \cite{finncher}, and we utilize it here. Accordingly to
\cite{sathyasim}, we have considered the
observational window confined to the frequency interval 
$40 - 750\,{\rm Hz}$. We suppose that the final frequency of inspiral 
is outside the considered bandwidth, so that the integral involved in
the definition of the inner product (\ref{eqn;innpr}) is evaluated 
on the same frequency range.

\subsection{Calculating the bounds}
\label{sunsec;calc}

In order to compare the bounds on parameter estimation errors
given by local
and global approaches, we computed 
the CRB, WWB and ZZB. The computational steps 
are described here, focusing particularly on 
the WWB and the ZZB. Our discussion is 
centered upon the evaluation of bounds for 
the time-of-arrival parameter, $t_a$ ($=\lambda_1$). Similar results
apply to $\Phi_a$ and $\tau$.

The CRB involves the computation of the diagonal
elements of the inverse of the Fisher information matrix (\ref{eqn;fimg}),
that is:
\be\label{eqn;crbh}
\epsilon_i^2 = {\cal J}^{-1}_{ii}\,;
\ee
for the signal (\ref{eqn;hf}) and the parameters (\ref{parchirp}),
we have
\be\label{eqn;derh}
\frac{\partial {\tilde h}}{\partial \lambda^j} = i\,
{\cal N} f^{-7/6} \psi_j(f) \exp\Psi(f)
\ee
and, therefore,
\be\label{eqn;fisherh}
{\cal J}_{jk} = \frac{\rho^2}{J}\int 
\psi_j(f)\psi_k(f)\frac{f^{-7/3}}{S_n(f)}\,df \,.
\ee
The evaluation of (\ref{eqn;crbh}) is straightforward from
Eqs. (\ref{eqn;psichirp}) and (\ref{eqn;fisherh}) and has been 
thoroughly studied in
many papers \cite{cf,pw,jk,kks,jkkt}, where further details can be found. 

The WWB involves the computation of Eq. (\ref{eqn;wwb4}) and therefore
of the $M\times N$ matrix ${\cal H}$ and the $N\times N$
matrix ${\cal G}$, where $M$ is the number of 
parameters (3 in this problem) and 
$N$ is the number of test points 
$\{\vdelta_j$; j=1, \ldots, N$\}$. The test points are 
now given explicitly by
$\Delta\lambda_j^{\nu}$ (the lower and upper indices labelling
the test points 
and the parameters, respectively). The elements of the matrix
${\cal H}$ are therefore ${\cal H}_{\nu j} = \Delta\lambda_j^{\nu}$. 

We will assume here  
that $\lambda^\nu$ has a uniform prior distribution with a  
region of support
that is much larger than the anticipated errors on the 
parameters. Therefore the prior will not impact on 
the calculation of the WWB. 

Eqs. (\ref{eqn;gmat}) and (\ref{eqn;mu}) read now
\be
{\cal G}_{jk} = 2
\frac{\exp[\rho^2\gamma(\Delta\lambda_j^{\nu}-\Delta\lambda_k^{\nu})/4]-
\exp[\rho^2\gamma(\Delta\lambda_j^{\nu}+\Delta\lambda_k^{\nu})/4]}
{\exp[\rho^2\gamma(\Delta\lambda_j^{\nu})/4] \,
\exp[\rho^2\gamma(\Delta\lambda_j^{\nu})/4]}\,,
\ee
and 
\be
\eta(1/2;\Delta\lambda_j^{\nu},\Delta\lambda_k^{\nu}) = \mu(1/2;
\Delta\lambda_j^{\nu}-\Delta\lambda_k^{\nu})\,.
\ee
A crucial issue for a reliable computation of the bound is
the placing of test points: (i) in order to get the
CRB in the limit of high SNR, the elements ${\cal H}_{\nu j}$
for $j=1,2,3$ (notice that $j$ runs from 1 to $N$) have been chosen
accordingly to ${\cal H}_{\nu j} = \delta_{\nu\alpha }
\Delta\lambda^{\alpha}_j$,
with $\Delta\lambda_j^{\alpha}\ll 1$ (here $\delta_{\nu \alpha}$ 
is the Kronecker
symbol); indeed, we always probe the primary peak of the ambiguity
function, as does the CRB; (ii) in order to get the tightest possible
bound at low SNR, we placed  the other test points (${\cal H}_{\nu j}$ for
$j > 3$) along the maxima of the ambiguity function, using the 
test-case problem presented in Sec. \ref{sec;ex} as a guideline.  
In Fig. \ref{fig;ambf} we show $\gamma(\Delta\lambda^{\nu})$, in the plane
$(\Delta t_a, \Delta\tau)$, maximized with respect to $\Delta \Phi_a$ 
(we already know that the regions where
the ambiguity function is small do not contribute significantly to the
result). The plot is enlightening as
$\gamma$ consists on a long, sharp ridge and clearly indicates that
the test points need to be spread along that curve. We placed 
up to 25 points (about the maximum permitted 
by the numerical routines implemented for the matrix inversion) with 
different choices of their separation and distance from the origin.
We noted, in fact, that with only 4 almost equally spaced points 
(spacing $\simeq 16\,{\rm ms}$), the result did not change significantly, 
in agreement with what found in the toy problem.
 
The evaluation of the ZZB involves the computation of the integral
(\ref{eqn;vzzb}), using the minimum probability error given by Eqs. 
(\ref{eqn;pmin}) and
(\ref{eqn;dist1}). As we have stressed before, the strategy of computation 
replaces here the spreading of test points with the selection of the
integration path. Our discussion of the evaluation of the WWB indicates
that the integration has to be performed 
along the ridge of the ambiguity function shown in Fig. \ref{fig;ambf}, 
in order to
produce the tightest bound. Of course, 
no ``valley filling'' function was needed,
as $\gamma$ is a smooth curve free from oscillations
for the signal that we were studying. 
We carried out the
integration up to a maximum displacement from the origin $\simeq 0.2\,
{\rm sec}$ (of the same order of the position of the last test point 
during the investigation of the WWB), after which no 
appreciable improvement was found. As in the case of the computation
of the WWB, the prior probability on $\lambda_1$ did not impact on the
bound, because the region of support was chosen to be
much larger than the anticipated
error. 

\subsection{Results}
\label{subsec:res}

The bounds calculated following the three different theoretical approaches 
(CRB, WWB and ZZB) were 
computed for values of $\rho$ in the relevant range
$7 \le \rho \le 25$. In the same SNR interval we compared these results
with those obtained by means of numerical simulations \cite{sathyasim} that 
implement maximum likelihood
estimators. The root-mean square error bounds on the 
time-of-arrival parameter $t_a$ are displayed as a function 
of $\rho$ in Fig. \ref{fig;ris}. All the bounds converge to the 
CRB at $\rho \sim 15$, and at this SNR the CRB is attained
by the maximum likelihood estimator. At smaller values of
SNR, the Bayesian bounds deviate from the CRB, providing
a slightly tighter result ($\simeq 6\%$ at $\rho = 10$ and 
$\simeq 25 \%$ at $\rho = 7$). This not too severe discrepancy can
be explained by the structure of $\gamma$:
both the local and global approaches
probe the ambiguity function around its origin, but the Bayesian 
bound is able to follow it further away from the origin
(see the discussion in Sec. \ref{sec;ex}).
The behavior of WWB and ZZB were found
to be very similar, although not exactly equal: the SNR threshold
at which they depart from the CRB is $\rho \simeq 14$, for the ZZB, and 
$\rho \simeq 12$, for the WWB, but the latter  
provides a better constraint at low SNRs. This agrees with results from other 
applications of the bound to time-of-arrival parameter 
estimation problems in radar and in sonar applications
(see \cite{bell} and reference therein). The striking feature of the
comparison is given by the maximum likelihood errors obtained in
numerical experiments: while matching the behaviour of (local and
global) lower 
bounds at high SNR ($\rho \simgt 15$), they produce errors that
are dramatically higher than expected theoretically at low SNRs: 
about $65\%$ 
at $\rho = 10$ and more than a factor of 2 at $\rho = 7$. 

\section{conclusions}

We have reassessed the issue of information extraction with respect to
observations of coalescing binaries by interferometric 
gravitational wave detectors. After discussing the main
properties of global and local bounds on parameter estimation,
we applied a set of these bounds to a gravitational wave 
parameter estimation problem. In particular we have 
introduced the Weiss-Weinstein and the Ziv-Zakai bounds. 
These provide
fundamental lower limits on the mean-square error on 
the parameters that describe a signal, independently of the 
actual estimator that is adopted in the data-analysis process. 
In addition these bounds easily incorporate any a priori information 
that is available about the problem and do not suffer from 
limitations that affect local bounds and the Bayesian 
version of the Cramer-Rao bound.  In short, these global bounds
can be used to benchmark the performance of any practical 
information extraction technique.

We have applied the bounds to the case
of laser interferometric
measurements of waveforms that are characteristic
of those emitted by inspiraling 
compact binaries in a background of noise that is
characteristic of the performance of first-stage detectors.  
Comparisons between the Cramer-Rao, 
Weiss-Weinstein and Ziv-Zakai bounds on the MSE
and actual maximum likelihood errors
obtained by numerical experiments, over a wide range of 
SNR, show that: (i) at high signal-to-noise
ratios (SNR $\simgt  15$) all the approaches
converge to the same value of the MSE. In this regime
one can regard the Fisher information matrix  
as a simple and reliable tool to 
compute ``realistic'' bounds on estimation errors. 
Maximum likelihood 
methods are probably adequate to extract astrophysical information 
from noisy data at these SNR's (although a 
definitive statement is premature,
pending a detailed analysis applied to more general 
waveforms); (ii) at low
signal-to-noise ratios (SNR $\simlt 10$, in which most of the events 
are likely to be recorded during the first years of operation of
the detectors) the WWB and ZZB produce
a more stringent constraint ($\simeq 25\%$ at $\rho = 7$)
on the MSE with respect to the CRB, indicating that the latter 
can underestimate the errors in this regime. Perhaps more 
seriously, all the bounds
are about two times smaller than the errors 
that are obtained in the numerical experiments. 

This analysis suggests that 
maximum likelihood techniques need to be refined, or
complemented, in order to
attain the lowest possible value of the errors. Our study
of toy-problems and the results from previous investigations in different
fields suggests that these are 
within a few percent of those predicted theoretically via the WWB and ZZB. 
A first attempt toward the understanding of 
the outcome of numerical experiments and the performances of
maximum likelihood estimators has been recently reported in \cite{BD}. 
We are currently exploring the
possibility of implementing conditional-mean estimators to improve
parameter 
estimation accuracy, and will report on this
in a forthcoming paper \cite{VN}. 
The conditional mean-estimator is generally intractable to implement 
as it requires the computation of a multi-dimensional integral over the
space of all the parameters (in realistic cases more than
ten), even though suitable strategies to reduce the amount of computation
have been proposed and successfully tested in (simple) cases
\cite{pasetti}. Hierarchical strategies combining maximum likelihood
and non-linear filtering could also speed up the process.

Finally, it is important to underscore the point that
the form in which we have presented the WWB
and ZZB  is completely general. It can be applied to 
parameter estimation problems for other kinds of 
signals ({\em e.g.} pulsars) and other instruments
and/or arrays of detectors. 

\acknowledgments
We are indebted to Kristine Bell for sending
us her thesis dissertation on the ZZB and graciously
answering many of our subsequent queries. In addition, we
thank Tony Weiss for fielding questions and providing
references to the WWB. We acknowledge helpful discussions
with Kostas Kokkotas, Andrzej Krolak, Bernard Schutz, and 
B.\,Sathyaprakash. AV acknowledges a fellowship from the Fondazione Della 
Riccia and the Department of Physics and Astronomy of 
the UWCC for the kind hospitality. The last months of AV's contribution
to this work were supported by the Max Planck Gesellschaft.

\newpage

\begin{figure}
\caption{
Schematic representation of the behaviour of the 
mean square error as a function
of the signal-to-noise ratio in a typical
non-linear estimation problem.
}
\label{fig;mse}
\end{figure}

\begin{figure}
\caption{
This diagram illustrates the link between the structure of the
ambiguity function and the accuracy of parameter estimation. 
The lower panel displays the ambiguity function for two 
different one-parameter signals, as a function of the displacement
$\Delta\theta$ from the true parameter value: $s_1$ (bold line) 
was chosen to have
an ambiguity function with one broad maximum; 
$s_2$ (dashed line) was designed to have a more structured
ambiguity function with many secondary side-lobes. For the special
signals considered here, both ambiguity functions
depend only on $\Delta\theta$, and not on the actual value $\theta$.  
In the upper panel we display the results of applying the
Cramer-Rao and the Weiss-Weinstein theory to bound  
the mean-square error $\epsilon_{\theta}$
in the estimation of the signal's parameter as a function of the
signal-to-noise ratio. See text for further details.
}
\label{fig;toy}
\end{figure}

\begin{figure}
\caption{
The ambiguity function, maximized with respect to
the phase-of-arrival $\Phi_a$, for the signal (\ref{eqn;hf}) as a function
of the time-of-arrival and of the chirp time.
}
\label{fig;ambf}
\end{figure}

\begin{figure}
\caption{
Comparison between local and global theoretical
bounds on the time-of-arrival error with actual 
parameter estimation errors obtained by applying 
the maximum likelihood method to simulated data
(bold line: CRB; unfilled circles: maximum of
WWB and ZZB;
full circles: maximum likelihood error).
}
\label{fig;ris}
\end{figure}

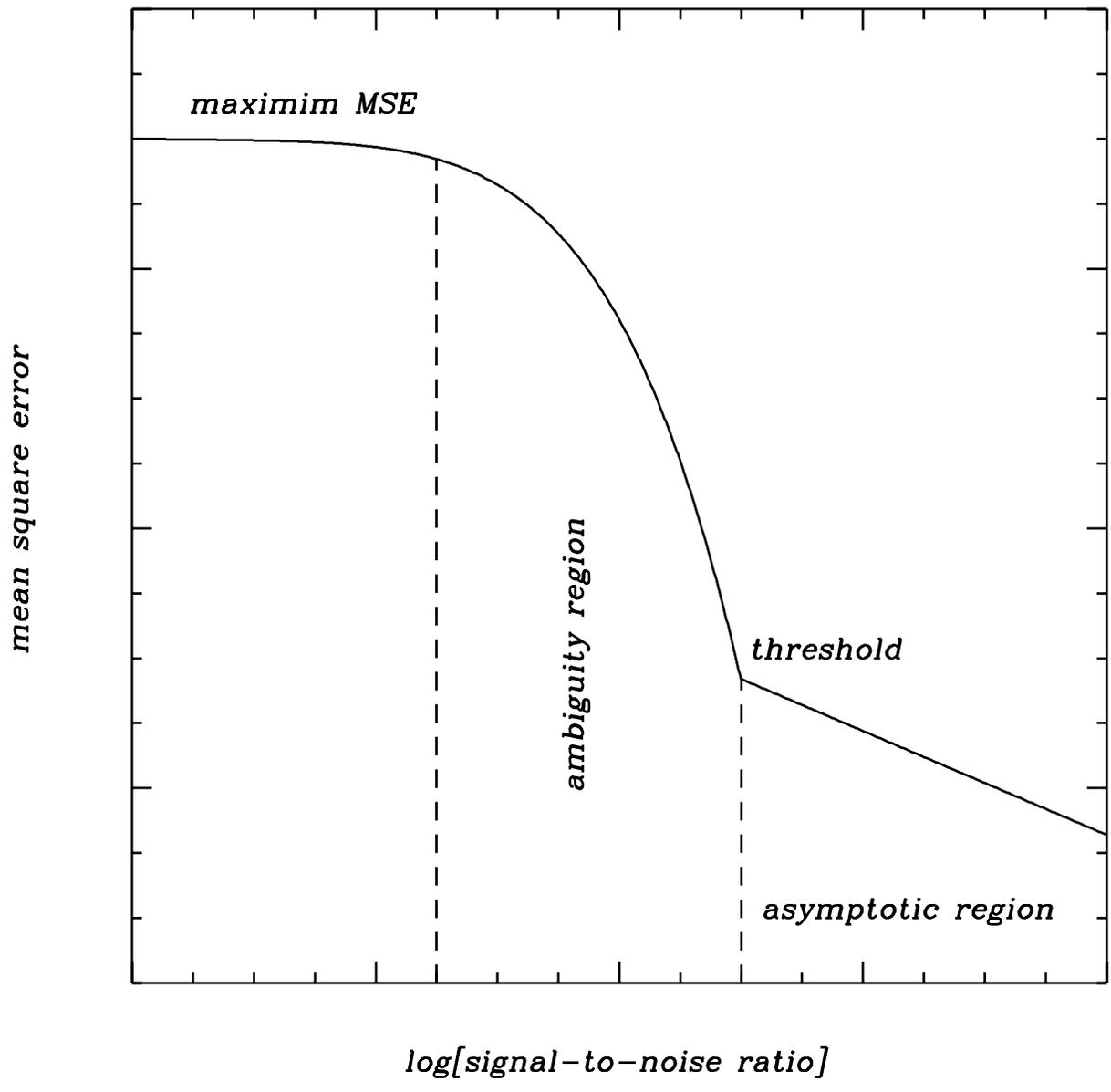
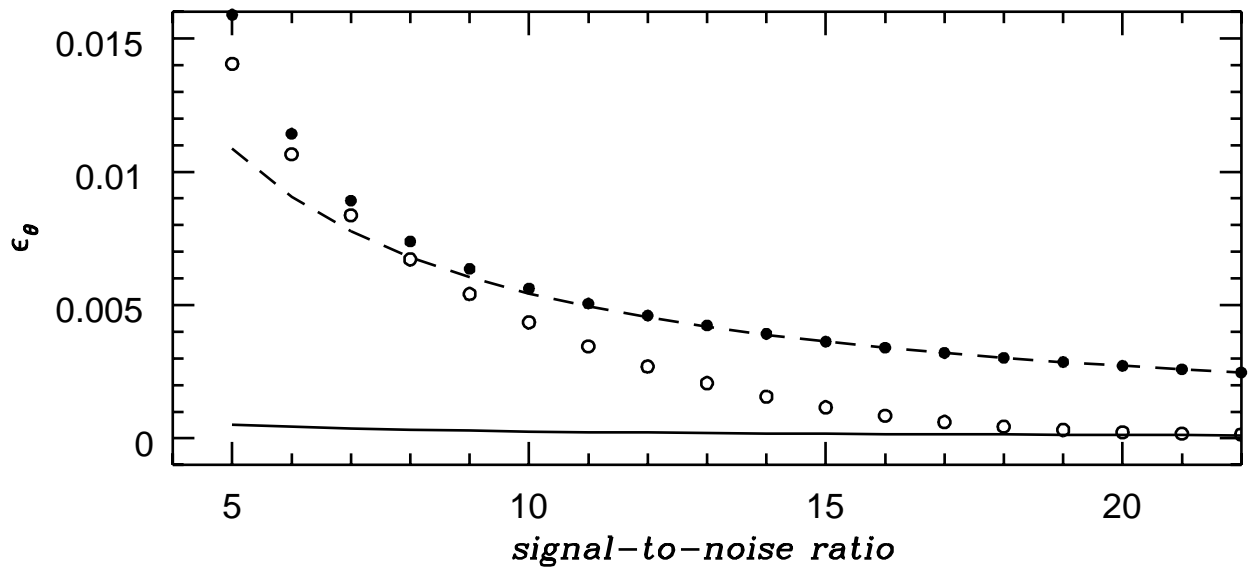
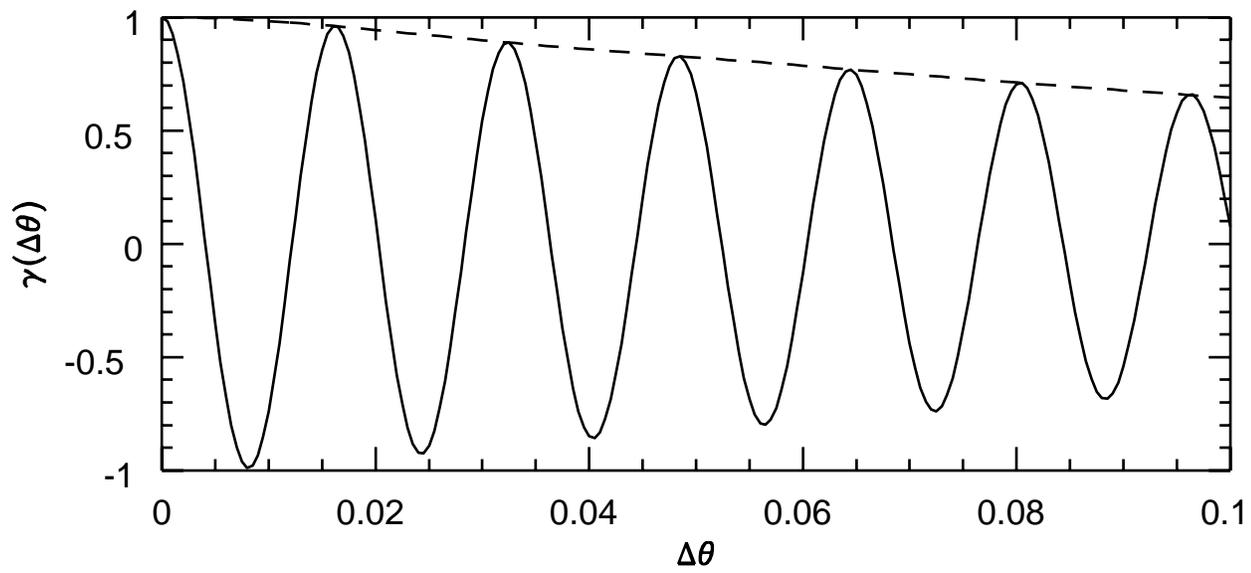
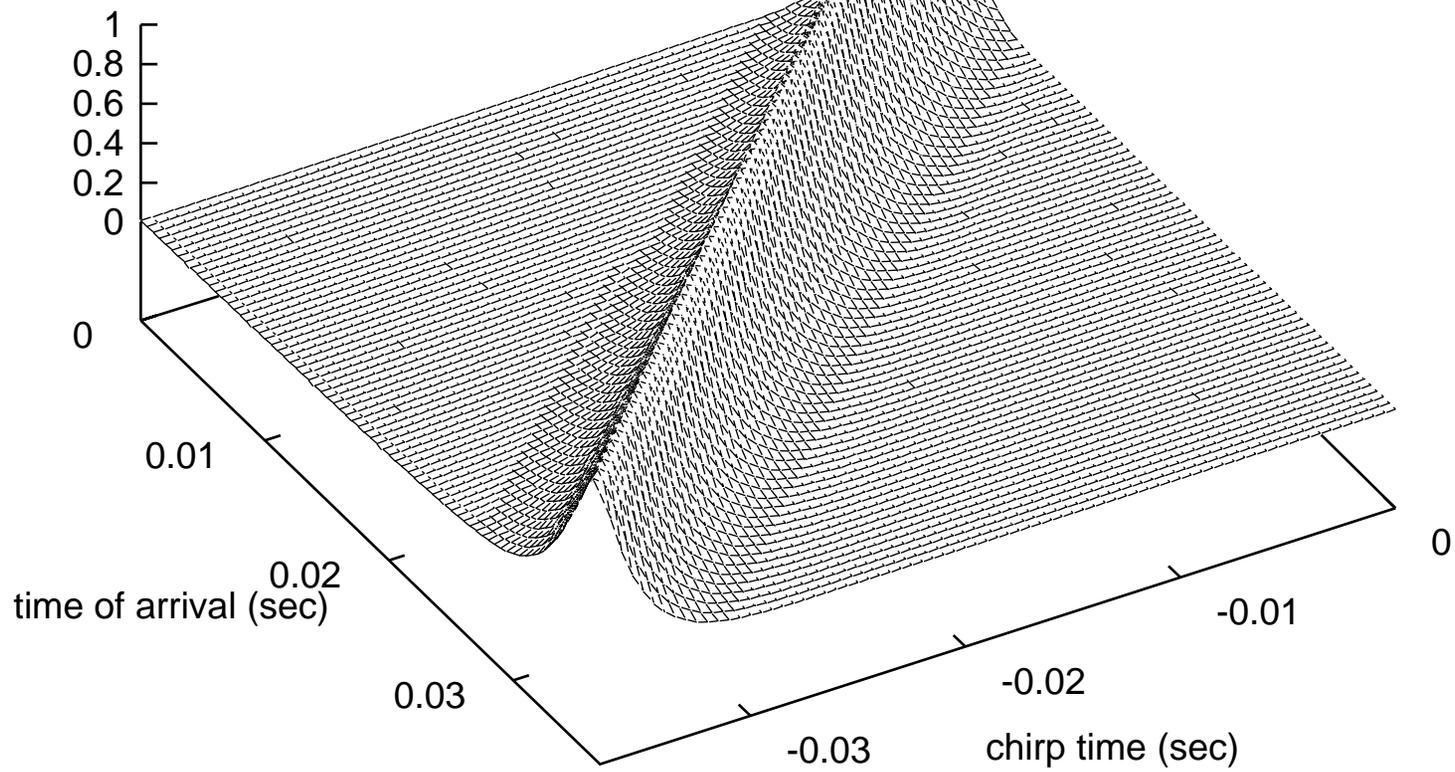
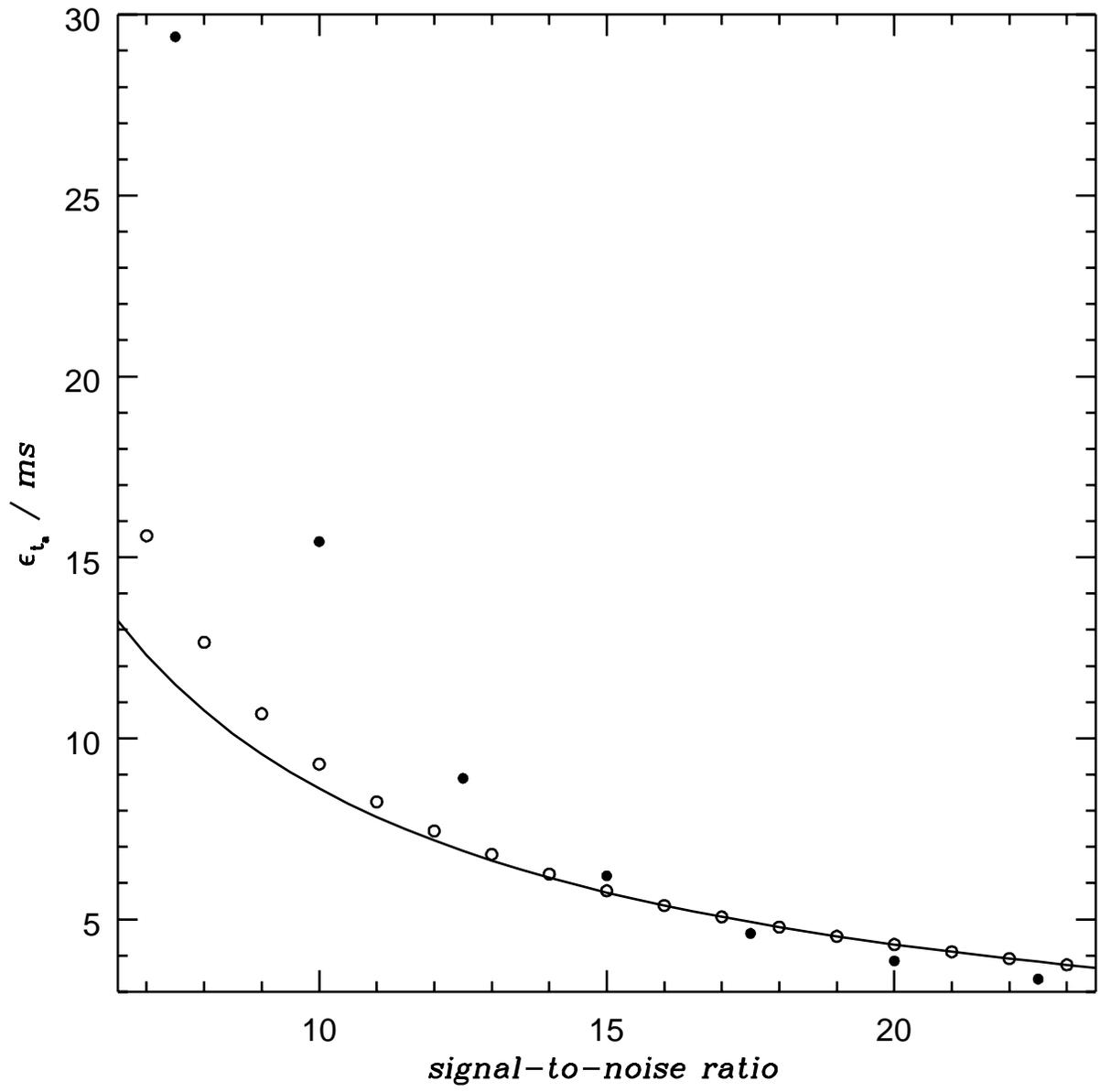
\end{document}